\documentclass{PoS}

\title{Quark flavour observables in 331 models \\in the flavour precision era}

\ShortTitle{Quark flavour observables in 331 models in the flavour precision era}

\author{Fulvia DE FAZIO\\
        INFN - Sezione di Bari\\
        E-mail: \email{fulvia.defazio@ba.infn.it}}

\abstract{I discuss a new Physics scenario, the 331 model, based on the gauge group $SU(3)_c \times SU(3)_L \times U(1)_X$. In particular, I elaborate on   correlations between flavour observables in the $B_d$ and $B_s$  systems that can help constraining the model parameters.}

\FullConference{The European Physical Society Conference on High Energy Physics \\
		 18-24 July, 2013\\
		 Stockholm, Sweden}

\newcommand{\be}{\begin{equation}}
\newcommand{\ee}{\end{equation}}
\newcommand{\bea}{\begin{eqnarray}}
\newcommand{\eea}{\end{eqnarray}}
\newcommand{\nn}{\nonumber}

\begin{document}

\section{Motivations}
Small tensions exist between the Standard Model (SM)  predictions and data on flavour observables. 
Understanding whether they represent   signals of new Physics (NP) requires a strong reduction in the uncertainty affecting such observables, both from the theoretical and experimental point of view.
Among such tensions, one can mention the  anomalies observed in the angular analysis of the decay $B \to K^* \mu^+ \mu^-$ \cite{Aaij:2013qta} as well as unexpectedly large branching fractions of semileptonic $B$ decays  to  a $\tau$ lepton in the final state. While improved 
measurements of the branching ratio of 
 the purely leptonic $B \to \tau \bar \nu_\tau$ mode  \cite{pdg} are close to theory predictions, a discrepancy between experimental data \cite{Lees:2012xj} and theory     holds in the case of the processes $B \to D^{(*)} \tau \bar \nu_\tau$, which seem to be anomalously enhanced with respect to SM predictions \cite{tau}.
NP models that might explain the anomaly without contributing to $B(B \to \tau \bar \nu_\tau)$ have been proposed \cite{tauNP} and await improved experimental data to be contrasted to.

 A intriguing question concerns the value of $|V_{ub}|$, since   the determination from exclusive $B$ decays turns out to be smaller than the one  from inclusive modes. 
Denoting by  scenario 1 (S1) that in which $|V_{ub}|$ assumes the smaller value $|V_{ub}|=3.1 \,\,10^{-3}$ and scenario 2 (S2) that in which $|V_{ub}|=4.0\,\,10^{-3}$, there are several observables  for which the agreement of the SM prediction   with data depends whether   S1 or S2 are realized.
In particular,   the following conclusions can be drawn in S1:
\begin{itemize}
\item S1 requires NP enhancing of  $B(B \to \tau \nu_\tau)$;
\item it reproduces the experimental value for the CP asymmetry $S_{J/\psi K_s}$ in $B \to J/\psi K_s$;
\item it suppresses $\epsilon_K$ with respect to experiment.
\end{itemize}
The opposite conclusions are reached in S2.
As for the mass differences $\Delta M_{d,s}$ in the $B_{d,s}-{\bar B}_{d,s}$ systems, their dependence on $|V_{ub}|$ is  mild, so that in both cases agreement with experiment is found within the uncertainties, even though NP models that predict a small suppression of these quantities are slightly favoured.

This discussion shows that 
 in order to understand whether in the LHC era a number of NP scenarios may be discarded and  the space of parameters of others can be constrained, more precise data and theoretical inputs are required. In view of this, it is interesting to try to predict what will happen in the flavour precision era, ahead of us, in which we can   assume that
\begin{itemize}
\item experimental data are affected by a much reduced uncertainty;
\item non perturbative parameters have been precisely calculated;
\item CKM matrix elements have been determined by means of tree-level decays (except for $|V_{ub}|$).
\end{itemize}
In this paper, following \cite{Buras:2012dp}, I discuss a NP model, the so-called 331 model, assuming that this era is already realized, showing
 the predicions for a number of flavour observables in $B_{d,s}$ systems.

\section{The model}
The name 331 encompasses a class of models based on the gauge group $SU(3)_c \times SU(3)_L \times U(1)_X$ \cite{Pisano:1991ee}, that is at first spontaneously broken to the Standard Model   group $SU(3)_c \times SU(2)_L \times U(1)_Y$ and then  undergoes the spontaneous symmetry breaking to $SU(3)_c \times U(1)_Q$. The extension of the gauge group with respect to SM leads to interesting consequences.
The first one is that the requirement of anomaly cancelation together with that of asymptotic freedom of QCD implies that the number of generations must necessarily be equal to the number of colours, hence giving an explanation for the existence of three generations.
Furthermore, quark generations should transform differently under the action of $SU(3)_L$. In particular, two quark generations  should transform as triplets, one as an antitriplet. Choosing the latter to be the third generation, this different treatment could be at the origin of the large top mass.

A fundamental relation holds among some of the generators of the group:
$Q=T_3+\beta T_8+X$ where $Q$ indicates the electric charge, $T_3$ and $T_8$ are two of the $SU(3)$ generators and $X$ is the generator of $U(1)_X$. $\beta$ is a key parameter that defines a specific variant of the model. Here I focus on the case $\beta=\frac{1}{\sqrt{3}}$ ($\overline{331} $ model), since the resulting scenario turns out to be phenomenologically more interesting than other variants. Moreover, the new gauge bosons, that are present due to the enlarged gauge group, have integer charges for this value of $\beta$. 

 The model comprises  several new particles. There are new gauge bosons $Y$ and $V$, whose charges depend on the considered variant. In  $\overline{331} $ they are a singly charged $Y^\pm$ boson and a neutral one $V^0 (\bar V^0)$. In all the variants a new neutral gauge boson $Z^\prime$ is present. This represents a very appealing feature, since $Z^\prime$ mediates tree level flavour changing neutral currents (FCNC) in the quark sector (couplings to leptons are instead universal). 
An extended Higgs sector is also present, with three $SU(3)_L$ triplets and one sextet.
Finally, new heavy fermions are predicted; I will not consider them  in this discussion.

As in the SM, quark mass eigenstates are defined upon rotation of flavour eigenstates through two unitary matrices $U_L$ (for up-type quarks) and $V_L$ (for down-type ones). The relation $V_{CKM}=U_L^\dagger V_L$ holds in analogy with the SM case. However, while in  SM $V_{CKM}$ appears only in charged current interactions and the two rotation matrices never appear individually, in this model only one matrix  between $U_L$ and $V_L$ can be expressed in terms of $V_{CKM}$ and the other one; the remaining rotation matrix enters in the $Z^\prime$ couplings to quarks. One can choose $V_L$ to be the surviving rotation matrix and  parametrize it as follows:
 \begin{equation}
V_L=\left(\begin{array}{ccc}
{\tilde c}_{12}{\tilde c}_{13} & {\tilde s}_{12}{\tilde c}_{23} e^{i \delta_3}-{\tilde c}_{12} {\tilde s}_{13} {\tilde s}_{23}e^{i(\delta_1
-\delta_2)} & {\tilde c}_{12}{\tilde c}_{23} {\tilde s}_{13} e^{i \delta_1}+ {\tilde s}_{12} {\tilde s}_{23}e^{i(\delta_2+\delta_3)} \\
-{\tilde c}_{13} {\tilde s}_{12}e^{-i\delta_3} & {\tilde c}_{12}{\tilde c}_{23} + {\tilde s}_{12}
 {\tilde s}_{13} {\tilde s}_{23}e^{i(\delta_1-\delta_2-\delta_3)} & - {\tilde s}_{12} {\tilde s}_{13}{\tilde c}_{23}e^{i(\delta_1 -\delta_3)}
-{\tilde c}_{12} {\tilde s}_{23} e^{i \delta_2} \\
- {\tilde s}_{13}e^{-i\delta_1} & -{\tilde c}_{13} {\tilde s}_{23}e^{-i\delta_2} & {\tilde c}_{13}{\tilde c}_{23}
\end{array}\right) \,\,.\label{VL-param}
\end{equation}
With this parametrization, considering  the Feynmann rules for $Z^\prime$ couplings to quarks,
it can be noticed  that the $B_d$ system involves only the parameters ${\tilde s}_{13}$ and $\delta_1$ while the $B_s$ system depends  on 
${\tilde s}_{23}$ and $\delta_2$. Stringent correlations between observables in $B_{d,s}$ sectors and in the kaon sector are found since kaon physics depends on ${\tilde s}_{13}$, ${\tilde s}_{23}$ and $\delta_2 - \delta_1$.
I refer to \cite{Buras:2012dp} for the analysis of these correlations and for predictions on kaon observables; in the next section I analyse the $B_{d,s}$ phenomenology, exploiting data on $\Delta F=2$ processes to constrain the above parameters in restricted oases and to predict correlations between $\Delta F=1$ observables that might allow to identify the right oasis, if any exists, which would mean that the model has a chance to be realized in Nature.

\section{Determining the optimal oasis in the parameter space}
As a preliminary, I fix the mass of the  $Z^\prime$ in the range $1 \le M_{Z^\prime} \le 3$ TeV,  in the reach of LHC. 
FCNC mediated by $Z^\prime$ involve only left-handed quarks and have the structure
\begin{equation}
i {\cal L}_L(Z^\prime)= i \left[\Delta_L^{sd}(Z^\prime)({\bar s} \gamma^\mu P_L d) +\Delta_L^{bd}(Z^\prime)({\bar b} \gamma^\mu P_L d)+ \Delta_L^{bs}(Z^\prime)({\bar b} \gamma^\mu P_L s)\right] Z^\prime_\mu \label{fcnc} \,\,,
\end{equation}
where $P_L=\frac{1-\gamma_5}{2}$ and the effective couplings $\Delta$ depend on the parameters ${\tilde s}_{13}$, ${\tilde s}_{23}$ and $\delta_1,\,\delta_2$.

I consder $\Delta F=2$ observables, namely the $B_{d,s} - {\bar B}_{d,s}$ mass differences $\Delta M_{d,s}$, the CP asymmetries  $S_{J/\psi K_s}$ in the decay $B_d \to J/\psi K_s$ and  $S_{J/\psi \phi}$ in the mode $B_s \to J/\psi \phi$. Imposing that $\Delta M_{d,s}$ vary in a range within $\pm 5\%$ of their experimental central value, while $S_{J/\psi K_s}$ and  $S_{J/\psi \phi}$ vary within a $2\sigma$ range of their experimental measurements \cite{Amhis:2012bh}, the resulting   constraints
\bea
0.48 \, {\rm ps}^{-1} \le &\Delta M_d& \le 0.53 \, {\rm ps}^{-1} \nn \\
0.64 \le & S_{J/\psi K_s} & \le 0.72 \nn \\
16.9 \, {\rm ps}^{-1} \le &\Delta M_s& \le 18.7 \, {\rm ps}^{-1} \nn \\
-0.15\le & S_{\psi \phi} & \le 0.15 \label{expDF2}
\eea
permit  to find allowed oases for the four parameters under consideration.
\begin{figure}[!h]
 \centering
\includegraphics[width = 0.45\textwidth]{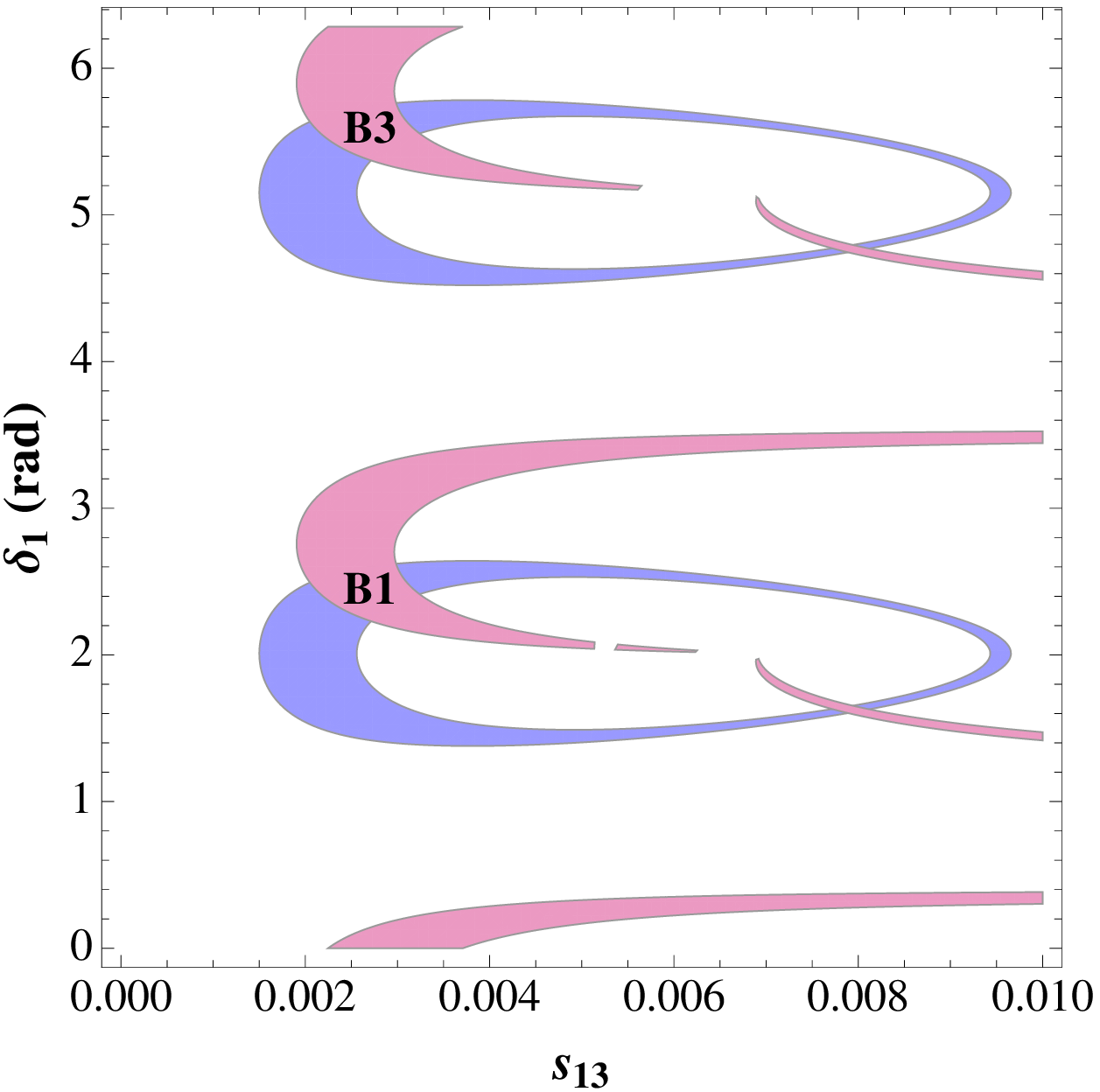} 
\includegraphics[width = 0.45\textwidth]{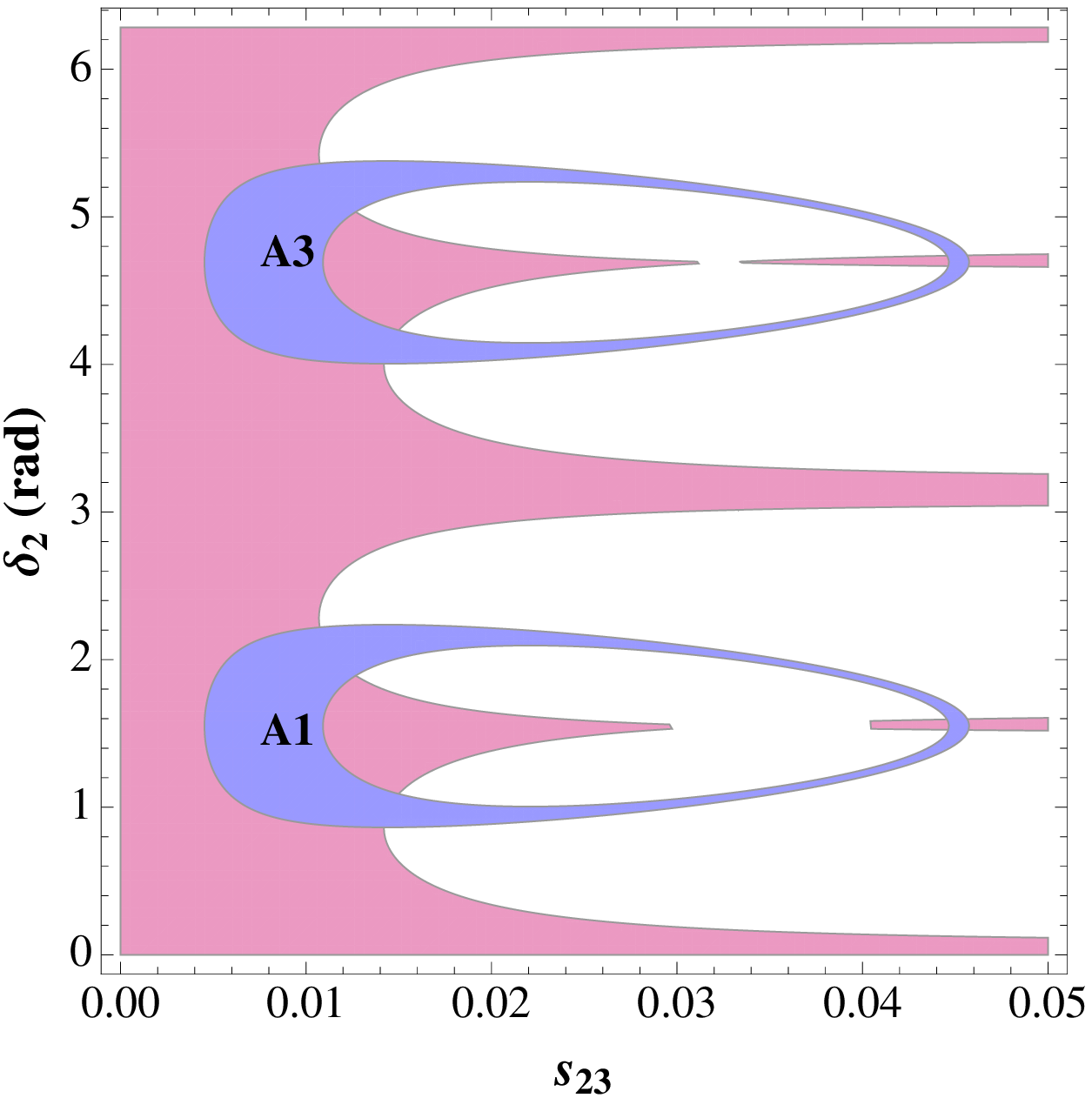} 
\caption{Left panel: Ranges for $\Delta M_d$ (violet region) and $S_{\psi K_s}$ (pink region). Right panel: Ranges for $\Delta M_s$ (violet region) and $S_{\psi \phi}$ (pink region).}\label{fig:oases}
\end{figure}
The result is shown in fig.\ref{fig:oases} in which the left panel refers to the $B_d$ case, the right one to the $B_s$ case, both in S1 scenario. Four oases are found in each case constraining the pairs $({\tilde s}_{13},\delta_1)$ (from $B_d$) and $({\tilde s}_{23},\delta_2)$ (from $B_s$). 
In both cases two large oases (A1, A3 for $B_s$, B1, B3 for $B_d$,  indicated in fig.\ref{fig:oases}) are present together with two small ones;  the latter can be discarded by imposing further experimental constraints on the mixing phase. 
Therefore I am going to discuss how to find the optimal oasis for the parameters $({\tilde s}_{13},\delta_1),({\tilde s}_{23},\delta_2)$ among the four pairs $(A_1,\,B_1)$, $(A_1,\,B_3)$, $(A_3,\,B_1)$, $(A_3,\,B_3)$.

For this purpose, other observables should be considered. 
Important modes  are $B_{s,d} \to \mu^+ \mu^-$ \cite{Buras:2013uqa}, that have been recently observed by the LHCb and CMS Collaborations \cite{Aaij:2013aka,Chatrchyan:2013bka}. The experimental analysis was optimized for $B_s$ case, and the result is in agreement with the SM. In the case of $B_d$ the SM prediction is  below the data, but the analysis still needs to be optimized for $B_d$ before conclusions can be reached.  

From the theory point of view, the SM effective hamiltonian for these decays depends only on a single  real function $Y_0(x_t)$ ($x_t=\frac{m_{top}^2}{M_W^2}$), which is  independent of the decaying meson and of lepton flavour. Its expression can be found e.g. in \cite{Buras:1998raa}.
The new $Z^\prime$ contribution modifies this function to a new one $Y(B_q)$ that now is different for $B_d$ and $B_s$ and has a complex phase $\theta_Y^{B_q}$, $q=d,s$  \cite{Buras:2012dp} . As a consequence, a CP asymmetry can be predicted in these modes, which reads: $S_{\mu^+ \mu^-}^q=\sin (2 \theta_Y^{B_q}- 2 \phi_{B_q})$, $\phi_{B_q}$ being a new phase entering in the mixing $B_q - {\bar B}_q$ that is absent in  SM. Therefore, these modes provide  two observables: their branching ratio and the CP asymmetry.
I discuss as an example the case of  $B_s$ system and show how the correlation between these observables and those related to $\Delta F=2$ processes can uniquely identify the optimal oasis. 
\begin{figure}[!h]
 \centering
\includegraphics[width = 0.49\textwidth]{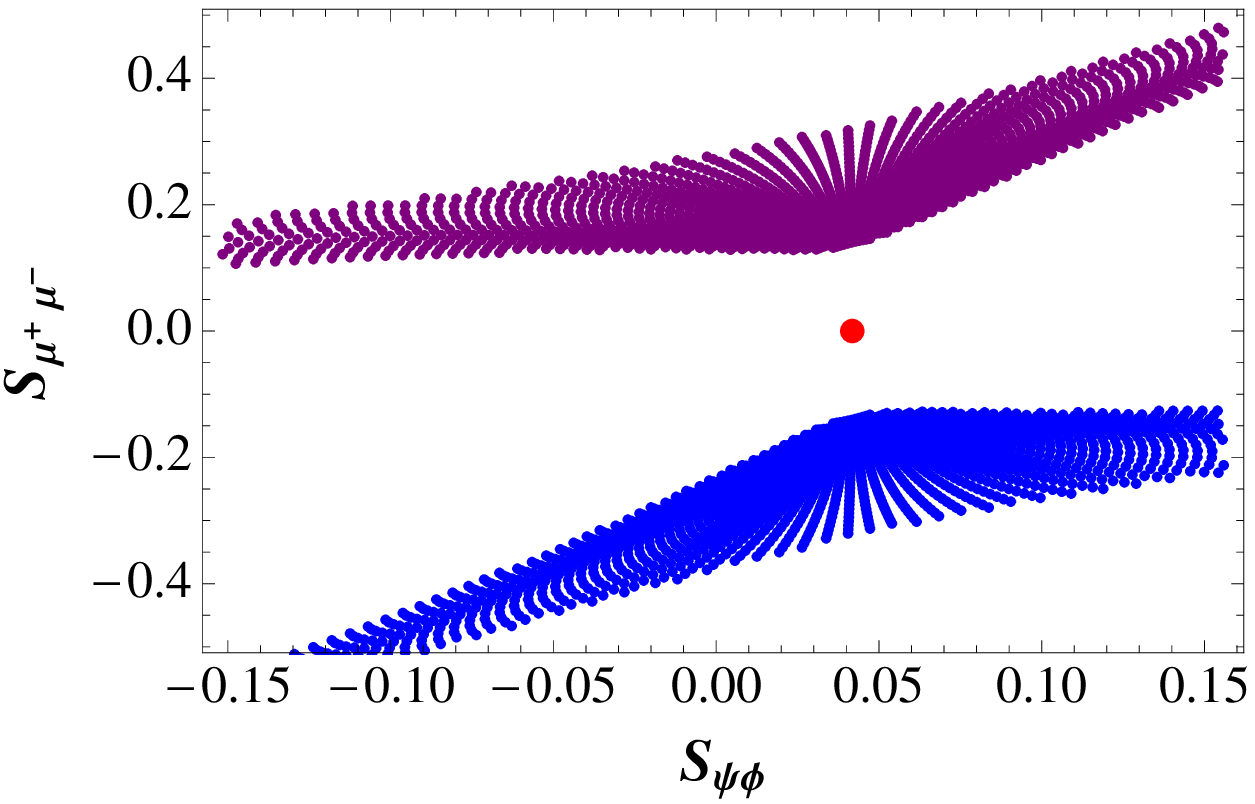} 
\includegraphics[width = 0.46\textwidth]{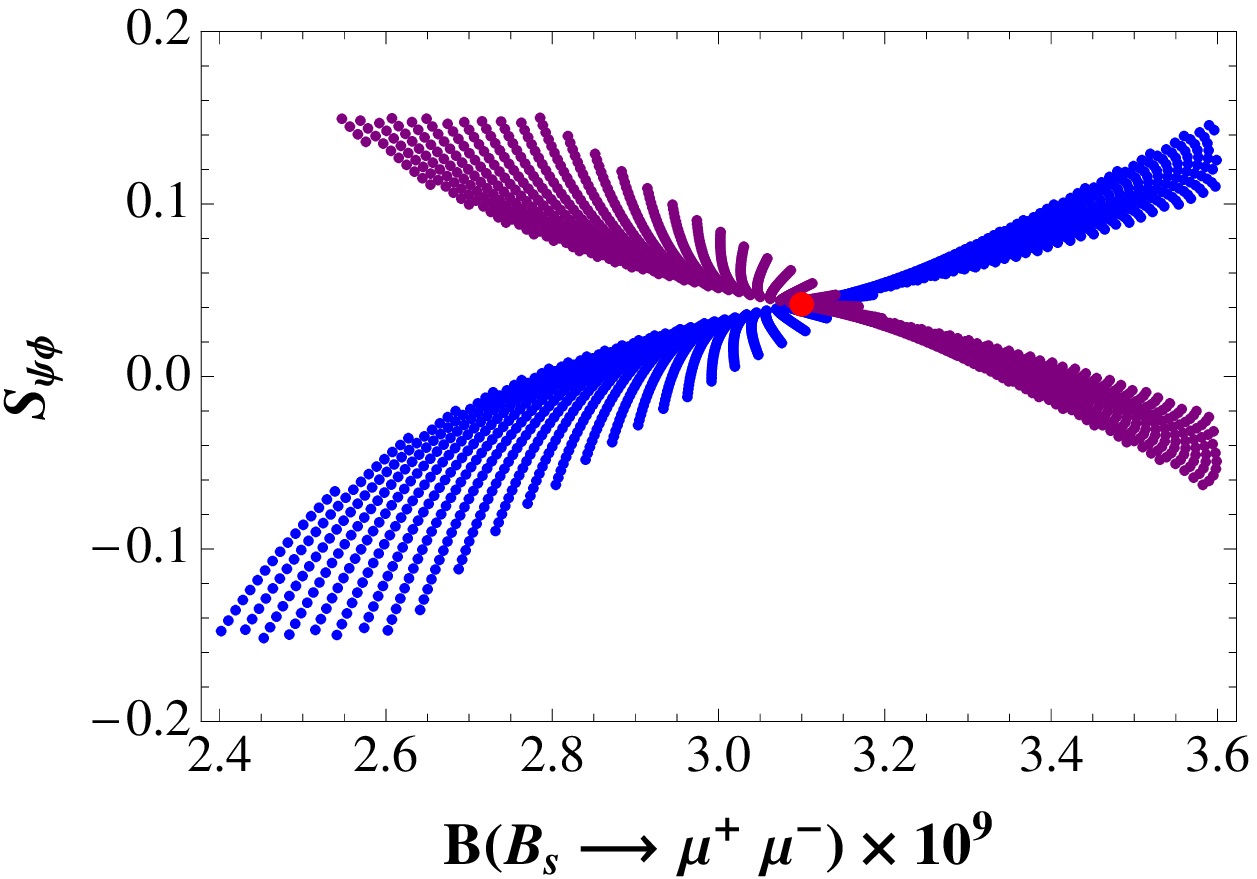} 
\caption{Left panel: $S_{\mu^+ \mu^-}^s$ vs $S_{\psi \phi}$. Right panel: $S_{\mu^+ \mu^-}^s$ vs $BR(B_s \to \mu^+ \mu^-)$. Blue regions correspond to the contribution of oasis A1, purple regions to the contribution of A3. The red points represent the SM predictions.}\label{fig:corr}
\end{figure}
Indeed, from fig.\ref{fig:corr} one can see a triple correlation. The plot in the left panel shows the two CP asymmetries $S_{\mu^+ \mu^-}^s$ versus $S_{\psi \phi}$. Measuring $S_{\mu^+ \mu^-}^s$ above its SM value, represented by the red point, would select the oasis A3, while in the opposite case the oasis A1 should be chosen. Once this has been done, the right  panel can be considered as a test of the model. In fact, it shows that $S_{\psi \phi}$ and the branching fraction $B(B_s \to \mu^+ \mu^-)$ are correlated in oasis A1 and anticorrelated in A3.
This means that if $A_1$ has been selected, $S_{\psi \phi}$ above (below) its SM value would imply $B(B_s \to \mu^+ \mu^-)$ also above (below) its SM value, while the opposite correlation occurs in $A_3$.  Measured incoherence between these two plots would mean that the model has to be discarded.

Many other observables can be considered,  that can help further in the search for the optimal oasis; in \cite{Buras:2012dp} a comprehensive analysis of $B_{d,s}$ and $K$ phenomenology can be found. It is worth  mentioning that the model can produce values of $\epsilon_K$ in agreeement with experiment both in S1 and S2 scenarios, i.e. independently of the value of $|V_{ub}|$.

\section{Conclusions}
331 models are interesting extensions of the SM, with dominant new Physics contributions  from tree level $Z^\prime$ exchanges. In this framework it is possible to remove existing tensions between the SM and experimental data when the mass of $Z^\prime$ is varied in the range $1 \le M({Z^\prime}) \le 3 $ TeV. In this case, the parameters of the model can be constrained in restricted oases using data on $\Delta F=2$ observables. Correlations among other quantities of interest, in particular processes induced by $\Delta F=1$ transitions, might allow to identify the optimal oasis and, simultaneously, serve as test of the model. I have shown this in a  particular case, i.e. underlying a triple correlation existing in the $B_s$ sector. 

If the mass of $Z^\prime$ is increased, it is still possible to find allowed oasis. However, the deviations from SM become smaller, so that the model could be hardly tested in this case.
The model provides also a concrete example of a scenario in which a new $Z^\prime$ boson is predicted to exist. A model independent analysis of correlations in the three systems of $B_d$, $B_s$ and $K$ mesons can be found in \cite{Buras:2012jb}.

{\bf Acknowledgments}
I thank  A.J. Buras, J. Girrbach and M.V. Carlucci for collaboration on the topics discussed in this talk, and  P. Colangelo for  comments.

\end{document}